\documentclass[aps,prl,twocolumn,floats,nofootinbib]{revtex4} 
\usepackage{graphics,graphicx,epsfig} 
\usepackage{amssymb,color} 
\usepackage{epsf,epstopdf,wrapfig} 
\usepackage {amsmath} 
 \usepackage{textgreek} 
\usepackage{sidecap}

\newcommand{\beq}{\begin{equation}} 
\newcommand{\eeq}{\end{equation}} 
\newcommand{\beqn}{\begin{eqnarray}} 
\newcommand{\eeqn}{\end{eqnarray}}

\begin{document} 

\title{Coarse--graining, fixed points, and scaling in a large population of neurons}

\author{Leenoy Meshulam,$^{1,2,3}$ Jeffrey L. Gauthier,$^{1}$ Carlos D. Brody,$^{1,4,5}$ David W. Tank,$^{1,2,4}$ and William Bialek$^{2,3,6}$} 

\affiliation{$^1$Princeton Neuroscience Institute, $^2$Joseph Henry Laboratories of Physics, $^3$Lewis--Sigler Institute for Integrative Genomics,  $^4$Department of Molecular Biology,  and $^5$Howard Hughes Medical Institute, Princeton University, Princeton, NJ 08544\\
	$^6$Initiative for the Theoretical Sciences, The Graduate Center, City University of New York, 365 Fifth Ave., New York, NY 10016} 

\begin{abstract} 
We develop a phenomenological coarse--graining procedure for activity in a large network of neurons, and apply this to  recordings from a population of 1000+ cells in the hippocampus.   Distributions of coarse--grained variables seem to approach a fixed non--Gaussian form, and we see evidence of scaling in both static and dynamic quantities.  These results suggest that the collective behavior of the network is described by a non--trivial fixed point.
\end{abstract}

\date{\today}

\maketitle

In systems with many degrees of freedom, it is natural to search for simplified, coarse--grained descriptions.  While this idea has a long history, our modern understanding is based on the renormalization group (RG).  In its conventional formulation, we start with the joint probability distribution for variables defined at the microscopic scale, and then coarse--grain by local averaging over small neighborhoods in space.  The joint distribution of coarse--grained variables evolves as we change the averaging scale, and in most cases the distribution becomes simpler as we move to larger scales.  Thus, macroscopic behaviors are simpler and more universal than their microscopic mechanisms \cite{kadanoff_66,wilson_75,wilson_79,cardy_96}.   Is it possible that simplification in the style of the RG will succeed in the more complex context of biological systems?

\begin{figure}[b]
	\includegraphics[width = \linewidth]{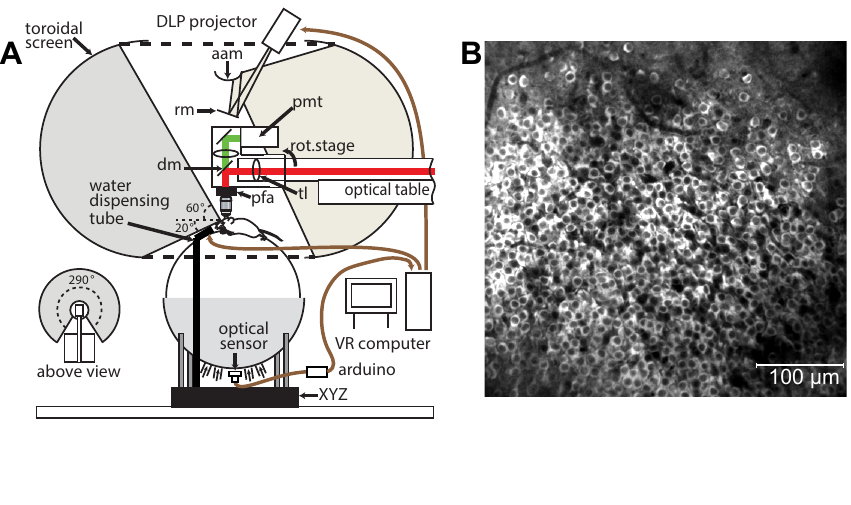}
\caption{(A) Schematic of the experiment, imaging inside the brain of a mouse running on a styrofoam ball. Motion of the ball advances the position of a virtual world projected on a surrounding toroidal screen. (B) Fluorescence image of neurons in the hippocampus expressing calcium sensitive fluorescent protein. 
\label{expt}}
\end{figure}

\begin{figure}
	\centering
	\includegraphics[width = \linewidth]{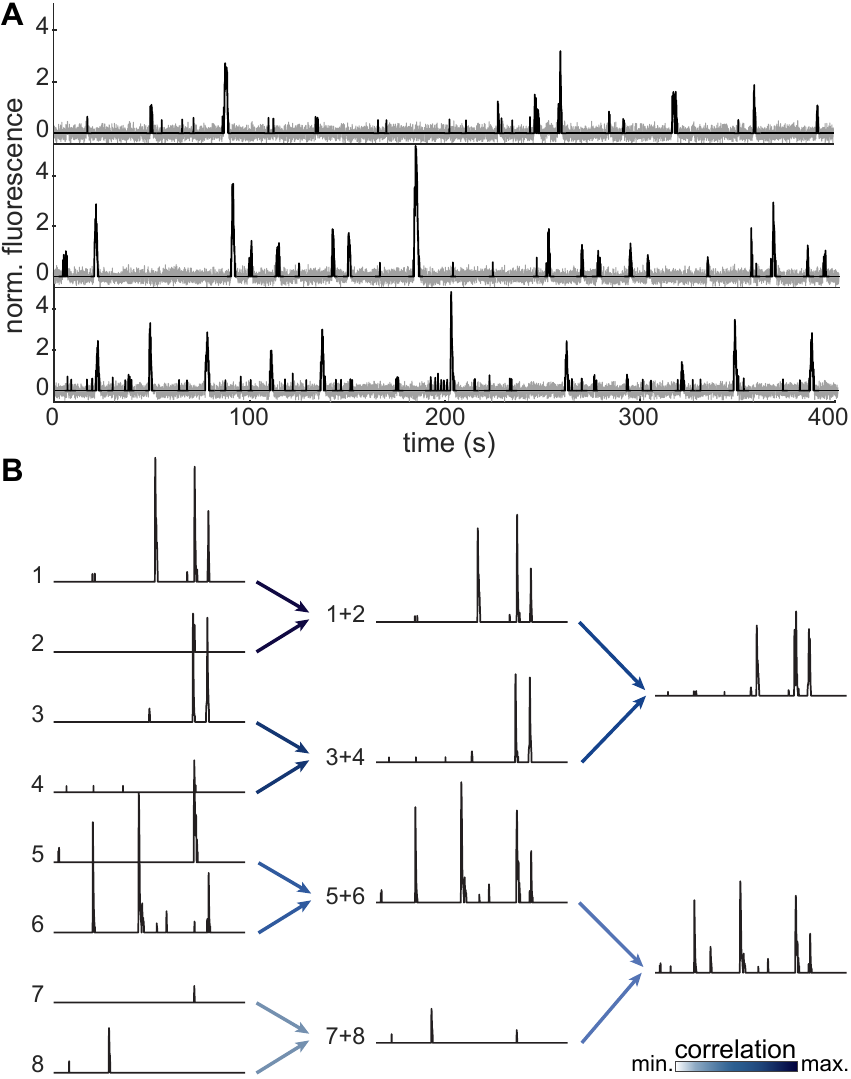}
	\caption{Fluorscence signals, denoising, and coarse--graining.  (A) Continuous fluorescence signals, raw in grey and denoised in black, for three neurons in our field of view.  Illustration of our coarse--graining procedure, for two iterations. (B) Activity of 8 example neurons. The first step in every iteration is to compute the hierarchy of correlations  between all pairs of neurons based on their activity. Then, maximally correlated pairs are grouped together by summing their activity, normalizing so the mean of nonzero values is one. Each cell can only participate in one pair, and all cells are grouped by the end of each iteration.  Darker arrows  correspond to stronger correlations in the pair.
		\label{schematic}}
\end{figure}

The  exploration of the brain has been revolutionized over the past decade by  methods to record, simultaneously, the electrical activity of large numbers of neurons \cite{segev+al_04,litke+al_04,harvey+al_09,marre+al_12,jun+al_17,chung+al_18,dombeck+al_10,harvey+al_12,ziv+al_13,nguyen+al_16,gauthier+tank_18}.  
Here we analyze experiments on 1000+ neurons in the CA1 region of the mouse hippocampus.  The mice are genetically engineered to express a protein whose fluorescence depends on the calcium concentration, which in turn follows electrical activity; fluorescence is measured with a scanning two--photon microscope as the mouse runs along a virtual linear track. Figure \ref{expt}A shows a schematic of the experiment, which has been described more fully elsewhere \cite{harvey+al_09,dombeck+al_10,gauthier+tank_18}. The field of view is $0.5\times0.5\,{\rm mm}^2$ (Fig  \ref{expt}B), and we identify 1485 cells. Fluorescence signals are sampled at 30 Hz, images are segmented to assign signals to individual neurons, and signals are denoised to reveal relatively infrequent transients above a background of silence (Fig \ref{schematic}A).   

In familiar applications of the RG,  microscopic variables have defined locations in space, and interactions are local, so it makes sense to average over spatial neighborhoods.  Neurons are extended objects, and make synaptic connections across distances comparable to our entire field of view, so locality is not a useful guide.   But in systems with local interactions, microscopic variables are most strongly correlated with the near spatial neighbors.  We will thus use correlation itself as a proxy for neighborhood.  We compute the correlation matrix of all the variables,  search greedily  for the most correlated pairs, and define a coarse--grained variable by the sum of the two microscopic variables in the pair \cite{significance}, as illustrated in Fig \ref{schematic}.  This can be iterated, placing the variables onto a binary tree; alternatively,  after $k$ iterations we have grouped the neurons into clusters of size $K = 2^k$, and each cluster is represented by a single coarse--grained variable \cite{mds}.

A technical point concerns the normalization of variables at each step of the coarse--graining.  We start with signals whose amplitude has an element of arbitrariness, being dependent on the relations between electrical activity and calcium concentration, and between calcium concentration and protein fluorescence.  Nonetheless, there are many moments in time when the signal is truly zero, representing the absence of activity.  We want to choose a normalization that removes the arbitrariness but preserves the meaning of zero, so we set the average amplitude of the nonzero signals in each cell equal to one, and restore this normalization at each step of coarse--graining.

Formally,  we start with  variables $\{x_{\rm i}(t)\}$ describing activity in each neuron ${\rm i}  = 1, 2, \cdots , N$ at time $t$; since our coarse--graining does not mix different moments in time, we drop this index for now.  We compute the correlations 
\begin{equation}
c_{\rm ij} = {{\langle\delta x_{\rm i}\delta x_{\rm j}\rangle}
\over
{\left[\langle(\delta x_{\rm i})^2\rangle\langle(\delta x_{\rm j})^2\rangle \right]^{1/2}}} ,
\label{cij_def}
\end{equation}
where $\delta x_{\rm i} = x_{\rm i} -\langle x_{\rm i} \rangle$.  We then search for the largest nondiagonal element of this matrix, identifying the maximally correlated pair ${\rm i}, {\rm j}_* ({\rm i})$, and construct the coarse--grained variable
\begin{equation}
x_{\rm i}^{(2)} = Z_{\rm i}^{(2)} \left( x_{\rm i} + x_{{\rm j}_* ({\rm i})} \right),
\end{equation}
where $Z_{\rm i}^{(2)}$ restores normalization as described above.  We remove the pair $[{\rm i}, {\rm j}_* ({\rm i})]$,  search for the next most correlated pair, and so on, greedily, until the original $N$ variables have become $\lfloor N/2\rfloor$ pairs.  We can iterate this process, generating $N_K = \lfloor N/K\rfloor$ clusters of size $K = 2^k$, represented by coarse--grained variables $\{x_{\rm i}^{(K)}\}$.

\begin{figure}[b]
\includegraphics[width=\linewidth]{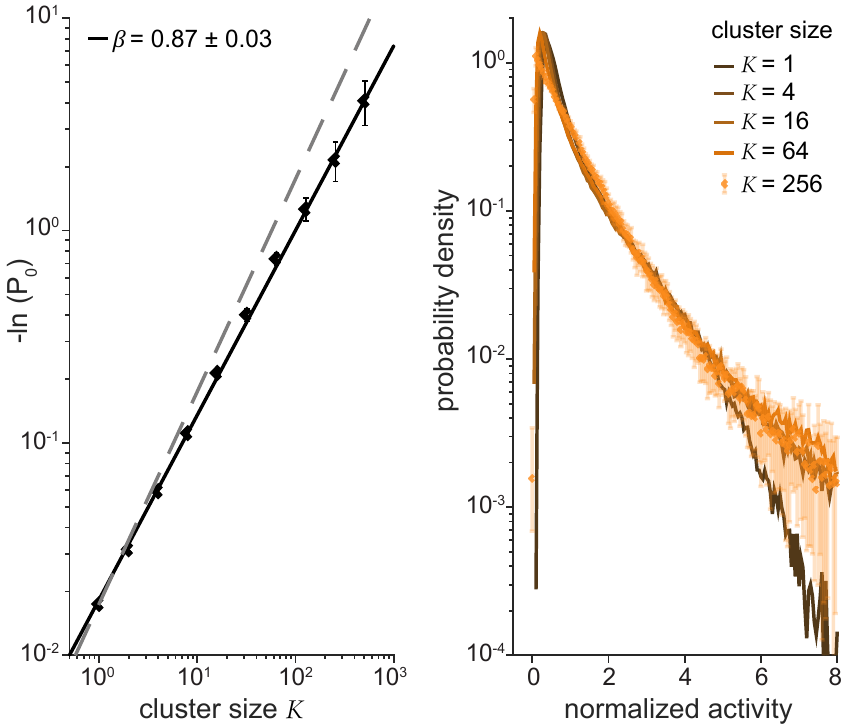}
\caption{Scaling in the probabilities of silence and activity. (left) Probability of silence as a function of cluster size. Dashed line is the expectation for independent neurons, and the solid line is from Eq (\ref{beta}). (right) Distribution of activity at different levels of coarse--graining, from Eq (\ref{defP}). Larger clusters corresponds to lighter colors.
\label{dists}}
\end{figure}

We would like to follow the joint distribution of variables at each step of coarse--graining, but this is impossible using only a finite set of samples \cite{clarify}.  Instead, as in the analysis of Monte Carlo simulations \cite{binder_81}, we follow the distribution of individual coarse--grained variables.  This distribution is a mixture of a delta function exactly at zero and a continuous density over positive values,
\begin{eqnarray}
P_K(x) 
&\equiv& {1\over {N_K}} \sum_{{\rm i}=1}^{N_K} {\bigg\langle} \delta\left( x - x_{\rm i}^{(K)}\right){\bigg\rangle}
\nonumber\\
&=& P_0(K) \delta (x) + [1-P_0(K)] Q_K(x) ,
\label{defP}
\end{eqnarray}
where our choice of normalization requires that 
\begin{equation}
\int_0^\infty dx\, Q_K(x) x  = 1.
\end{equation}
Figure \ref{dists} shows the behavior of $P_0(K)$ and $Q_K(x)$ from the microscopic scale $K=1$ out to $K=256$.

If the coarse--grained activity of a cluster is zero, all the microscopic variables in that cluster must be zero, so that $P_0(K)$ measures the probability of silence in clusters of size $K$.  If neurons were independent, then this probability would fall exponentially with $K$; in fact the data are well described by
\begin{equation}
P_0(K) = \exp(-a K^{\tilde\beta}),
\label{beta}
\end{equation}
with $\beta = 0.87\pm0.03$ \cite{errors}.    This scaling suggests that correlations among neurons  are self--similar \cite{scale_range}.

Coarse--graining replaces individual variables by averages over increasingly many microscopic variables.  If correlations among the microscopic variables are sufficiently weak, the central limit theorem will drive the distribution of coarse--grained variables toward a fixed Gaussian form;  a profound result of the RG is the existence of non--Gaussian fixed points.  This seems to be what is happening with $Q_K(x)$, as shown at right in Fig \ref{dists}.  There are surprisingly small changes in this distribution, even as $K$ varies by two orders of magnitude.  The changes that do occur are leading to a very simple, exponential form at large $K$, and there is no sign of approach to a Gaussian.

If correlations are self--similar, then we should be able to see this in more detail by looking inside the clusters of size $K$, which are analogous to  spatially contiguous regions in a system with local interactions.  We recall that, in systems with translation invariance, the matrix of correlations among microscopic variables is diagonalized by a Fourier transform, and that the eigenvalues $\lambda$ of the covariance matrix are the power spectrum or propagator $G(k)$.  At a fixed point of the RG this propagator will be scale invariant,  $\lambda = G(k) = A/k^{2-\eta}$, where the wavevector  $k$ indexes the eigenvalues from largest (at small $k$) to smallest (at large $k$), and in $d$ dimensions the eigenvalue at $k$ is of rank $\sim (Lk)^d$, where $L$ is the linear size of the system.  The number of variables in the system is $K \sim (L/a)^d$, where $a$ is the lattice spacing and the largest $k\sim 1/a$.  Putting these factors together we have
\begin{equation}
\lambda = B \left( {K\over{\rm rank}}\right)^\mu,
\label{def_mu}
\end{equation}
with $\mu = (2-\eta)/d$ and $B = Aa^{2-\eta}$.  Thus scale invariance implies  both  a power--law dependence of the eigenvalue on rank and  a dependence only on fractional rank ($ {\rm rank}/K$) when we compare systems of different sizes.

\begin{figure}[h]
	\centering
	\includegraphics[width=\linewidth]{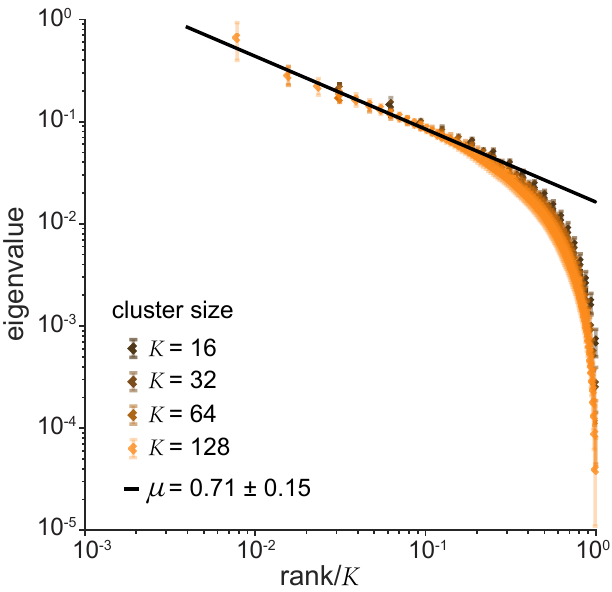}
	\caption{Scaling in eigenvalues of the covariance matrix spectra, $C_{\rm ij} =\langle\delta x_{\rm i}\delta x_{\rm j}\rangle$, for clusters of different sizes. Larger cluster corresponds to lighter color. Solid line is the fit to Eq (\ref{def_mu}).
		\label{eigs}}
\end{figure}

Figure \ref{eigs} shows the eigenvalues of the covariance matrix, $C_{\rm ij} =\langle\delta x_{\rm i}\delta x_{\rm j}\rangle$,
in clusters of size $K = 16, 32, 64, 128$.    A power--law dependence on rank is visible, albeit only over little more than one decade; perhaps more compelling is the dependence of the spectrum on relative rank,  accurate over much of the spectrum within the small error bars of our measurements.  The best fit exponent is $\mu = 0.71 \pm 0.15$.

\begin{figure}
\includegraphics[width=\linewidth]{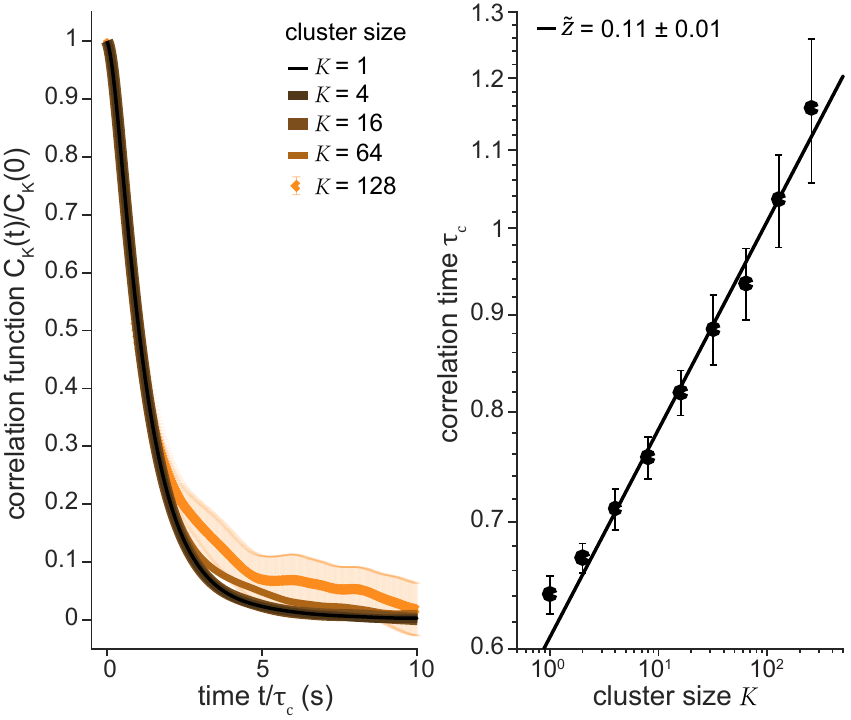}
\caption{Dynamic scaling. (left) Correlation functions for different cluster sizes [Eq (\ref{CK_def})].  We show $K =1, 4, 16, 64, 128$ (last one with error bars), where color lightens as $K$ increases, illustrating the scaling behavior when we measure time in units of $\tau_c(K)$.  (right) Dependence of correlation time on cluster size, with fit to Eq (\ref{tauc}).
\label{dynamics}}
\end{figure}

In systems with local interactions, the spread of correlations throughout the system takes time.  If we are near a fixed point of the RG,  then we will see dynamic scaling, with fluctuations on length scale $\ell$ relaxing on time scale $\tau \propto \ell^z$.  Although interactions in the neural network are not local, we have clustered neurons into blocks based on the strength of their correlations, and we might expect that larger blocks will relax more slowly.  To test this, we compute the temporal correlation functions for coarse--grained variables,
\begin{equation}
C_K(t) = {1\over {N_K}} \sum_{{\rm i}=1}^{N_K} \langle \delta x_{\rm i}^{(K)} (t_0) \delta x_{\rm i}^{(K)} (t_0 + t)\rangle .
\label{CK_def}
\end{equation}
Qualitatively, the decay of $C_K(t)$ is slower at larger $K$, but as we see  in Fig \ref{dynamics} the correlation functions at different $K$ have the same form within error bars if we scale the time axis by a correlation time $\tau_c(K)$.  Although the range of $\tau_c$ is small, we see that
\begin{equation}
\tau_c(K) = \tau_1 K^{\tilde z} ,
\label{tauc}
\end{equation}
except for the smallest $K$ where the dynamics are limited by the response time of the fluorescent indicator molecule itself.  Quantitatively, $\tilde z = 0.11 \pm 0.01$.

Before  interpreting these results, we make two observations which will be explored in detail elsewhere \cite{meshulam+al_18b}.  First, everything we have done here can be redone by first discretizing the continuous fluorescence signals into a binary on/off description of activity in single neurons, as in Ref \cite{meshulam+al_17}.  Again we see an approach to a fixed exponential rather than Gaussian distribution and power--law scaling; all exponents agree within error bars.    We have done the same experiment and analysis independently in three different mice; importantly there are no ``identified neurons'' in the mammalian brain, so we can revisit the same region of the hippocampus in another animal, but there is no sense in which we revisit the same network of neurons.  Nonetheless we see the same  approach to a fixed distribution and power--law scaling, with exponents again agreeing within error bars; this is true even for $\tilde\beta$, which has error bars in the second decimal place.  These results suggest, strongly, that behaviors we have identified via coarse--graining are independent of variations in microscopic detail, as we hope.

Second, we need to consider the relationship of our observations to the salient qualitative fact about the rodent hippocampus, namely that many of the neurons in this brain area are ``place cells'' \cite{okeefe+dostrovsky_71,okeefe+nadel}.  Place cells are active only when the animal visits a small, compact region of space, and are silent otherwise; together the activity in the place cell population is thought to form a cognitive map that guides the animal's navigation.    We find that the spatial localization of activity  is preserved by our coarse--graining procedure, although it was not designed specifically to do this.  In fact fewer than half of the cells in the population that we study are place cells in this particular environment, but after several steps of coarse--graining essentially all of the coarse--grained variables  have well developed place fields.  On the other hand, the scaling behavior that we see is not a simple consequence of place field structure.  To test this, we estimate for each cell the probability of being active at each position, and then simulate a population of cells that are active with this probability but independently of one another.  In smaller populations we know that this independent place cell model fails to capture important aspects of the correlation structure \cite{meshulam+al_17}, and here we find that it does not exhibit the scaling shown in Figs \ref{dists}--\ref{dynamics}.  

In equilibrium statistical mechanics problems with local interactions,  a fixed distribution and associated power--law scaling behaviors are signatures of a system poised near a critical point in its phase diagram.  The idea that networks of neurons might be near to criticality is intriguing, and has been discussed for more than a decade \cite{mora+bialek_11}.    One version of this idea focuses on ``avalanches'' of sequential activity in neurons \cite{beggs+plenz_03,friedman+al_12},   by analogy to what happens in the early sandpile models for self--organized criticality \cite{bak+al_87}.  A very different version focuses on the distribution over microscopic states in the network at a single instant of time  \cite{tkacik+al_06,tkacik+al_14}, and is  more closely connected to criticality in equilibrium statistical mechanics.    Related ideas have been explored in other biological systems, from biochemical and genetic networks \cite{socolar+kauffman_03,ramo+al_06,nykter+al_08,krotov+al_14} to flocks and swarms \cite{bialek+al_14,cavagna+al_17}.   In our modern view, invariance of probability distributions under iterated coarse--graining---a fixed point of the renormalization group---may be the most fundamental test for criticality, and has meaning independent of analogies to thermodynamics.

Although often thought of in connection with critical phenomena, a  fundamental result of the RG is the existence of irrelevant operators, which means that successive steps of coarse--graining lead to  simpler and more universal models.  Although the RG transformation begins by reducing the number of degrees of freedom in the system, simplification does not result from this dimensionality reduction but rather from the flow through the space of models.  The fact that our phenomenological approach to coarse--graining  gives results which are familiar from successful applications of the RG in statistical physics encourages us to think that simpler and more universal theories of neural network dynamics are possible.

\begin{acknowledgments}
We thank S Bradde, A Cavagna, DS Fisher, I Giardina, MO Magnasco, SE Palmer, and DJ Schwab for  helpful discussions.  Work supported in part by the National Science Foundation through the Center for the Physics of Biological Function (PHY--1734030), the Center for the Science of Information (CCF--0939370), and  PHY--1607612; by the Simons Collaboration on the Global Brain; and by the Howard Hughes Medical Institute.
\end{acknowledgments}


\begin{thebibliography}{99}
\bibitem{significance}
Since we are using the strongest correlations, our coarse--graining procedure is not very sensitive to the spurious correlations that arise from finite sample size.
\bibitem{clarify}
When we use the renormalization group to study models, we indeed follow the flow of the joint distribution, in various approximations \cite{wilson_75,cardy_96}.   When we are trying to analyze data, either from experiments or from simulations, this is not possible.
\bibitem{errors} 
Error bars for all scaling behaviors reported in this work were estimated as the standard deviation across quarters of the data. To respect temporal correlations, the location of the quarter was chosen at random, but time points remained in order inside it. For each quarter of the data, exponents are estimated as the slope of the best fit line on a log--log scale.
\bibitem{scale_range}
Scaling usually is an asymptotic behavior, but in Fig \ref{dists} we see consistency with a power--law across almost the full range from $K=1$ to $K=512$.  If we fit only for $K\geq 32$, for example, we find the same value of $\tilde\beta$ within error bars; there is no sign that these larger values of $K$ are more consistent with $\tilde\beta = 1$.  Thanks to DS Fisher for asking about this.
\end{thebibliography}
\end{document}